\documentclass[twocolumn,aps,prl,preprintnumbers,superscriptaddress]{revtex4}
\usepackage{}
\usepackage{bbm}
\usepackage[latin9]{inputenc}
\usepackage{amsmath}
\usepackage{amssymb}
\usepackage{graphicx}
\usepackage{mathrsfs}
\usepackage{amsfonts}
\usepackage{amsthm}
\usepackage{color}
\usepackage{txfonts}
\usepackage{lipsum}
\usepackage{gensymb}
\usepackage{ulem}
\usepackage[colorlinks=true,citecolor=blue,linkcolor=blue,urlcolor=blue,anchorcolor=blue]{hyperref}%
\hypersetup{colorlinks=true,citecolor=blue,linkcolor=blue,urlcolor=blue}
\providecommand{\U}[1]{\protect\rule{.1in}{.1in}}
\makeatletter

\newcommand{\Rmnum}[1]{\expandafter\@slowromancap\romannumeral #1@}
\makeatother
\makeatletter
\@ifundefined{textcolor}{}
{
\definecolor{BLACK}{gray}{0}
\definecolor{WHITE}{gray}{1}
\definecolor{RED}{rgb}{1,0,0}
\definecolor{GREEN}{rgb}{0,1,0}
\definecolor{BLUE}{rgb}{0,0,1}
\definecolor{CYAN}{cmyk}{1,0,0,0}
\definecolor{MAGENTA}{cmyk}{0,1,0,0}
\definecolor{YELLOW}{cmyk}{0,0,1,0}
}
\makeatother
\begin{document}
\title{Topological Magnon Frequency Combs}
\author{Zhixiong Li}
\affiliation{School of Physics, Central South University, Changsha 410083, China}
\author{Xuejuan Liu}
\affiliation{College of Physics and Electronic Engineering, Chengdu Normal University, Chengdu 611130, China}
\author{Zhejunyu Jin}
\affiliation{School of Physics and State Key Laboratory of Electronic Thin Films and Integrated Devices, University of Electronic Science and Technology of China, Chengdu 610054, China}
\author{Guanghua Guo}
\affiliation{School of Physics, Central South University, Changsha 410083, China}
\author{Xingen Zheng}
\email[Contact author: ]{zhengxingen@sslab.org.cn}
\affiliation{Songshan Lake Materials Laboratory, Dongguan 523808, China}
\author{Peng Yan}
\email[Contact author: ]{yan@uestc.edu.cn}
\affiliation{School of Physics and State Key Laboratory of Electronic Thin Films and Integrated Devices, University of Electronic Science and Technology of China, Chengdu 610054, China}

\begin{abstract}
Exploring the synergy between topological physics and nonlinear dynamics unveils profound insights into emergent states of matter. Inspired by recent experimental demonstrations of topological frequency combs in photonics, we theoretically introduce topological magnon frequency combs (MFCs) in a two-dimensional triangular skyrmion lattice. Computing the Chern numbers of magnon bands reveals robust chiral edge states. Strikingly, these topological MFCs originate from nonlinear four-magnon scattering among the chiral edge modes, activated by dual-frequency driving without an amplitude threshold. Comb spacings are readily tunable through excitation frequency detuning. Micromagnetic simulations validate our predictions with good concordance. This work paves the way for defect-immune magnonic devices exploiting MFCs and sparks investigations into topological-nonlinear phenomena in magnetic systems.
\end{abstract}

\maketitle
\textit{Introduction}---In optical systems, nonlinear effects can be leveraged to produce frequency combs, spectra consisting of discrete, equally spaced lines \cite{UdemN2002,DelHayeN2007}. This signature feature powers revolutionary applications, spanning atomic clocks \cite{Ludlow2015}, satellite navigation \cite{Lezius2016}, low-noise microwave generation, and optical spectroscopy \cite{Picque2019}. The extraordinary success of optical frequency combs has catalyzed analogous pursuits in varied arenas, including phononics \cite{CaoPRL2014,GanesanPRL2017} and mechanics \cite{CzaplewskiPRL2018}. Recently, Wang \textit{et al.} \cite{WangPRL2021,WangPRL2022} unveiled magnon frequency combs (MFCs) via nonlinear interactions between magnons and magnetic solitons, such as skyrmions and vortices. This advance has aroused widespread interest, driven by compelling applications in information processing and quantum computing \cite{Hula2022,Liu2023,zhou2021,Rao2023,JinPRL2023,YaoPRB2023,Xu2023PRL,WangNP2024,LiuPRB2024,Shennpj2024,ZhangAPL2024,LiangNL2024,LxjPRB2024}. Despite abundant research on MFCs across magnetic systems, dominant strategies depend on topologically trivial magnons, leaving them vulnerable to defects and disorder that undermine performance in practical materials. Meanwhile, topological magnonics \cite{WangJAP2021,McClartyARCMP2022,Libook2023,YuPR2024} has risen as a compelling paradigm for engineering resilient magnonic devices with cutting-edge functionalities, like magnon diodes, beam splitters, and interferometers \cite{WangPRA2018,ShindouPRB2013}. These harness topological magnon edge modes, localized at boundaries, chiral \cite{MookPRB2014,WangPRB2017}, and innately resistant to backscattering from disorder or perturbations. Yet, topological magnon inquiries have largely stayed within the linear realm, with nonlinear dynamics still vastly underexplored.
\begin{figure}[!htbp]
\begin{centering}
\includegraphics[width=0.44\textwidth]{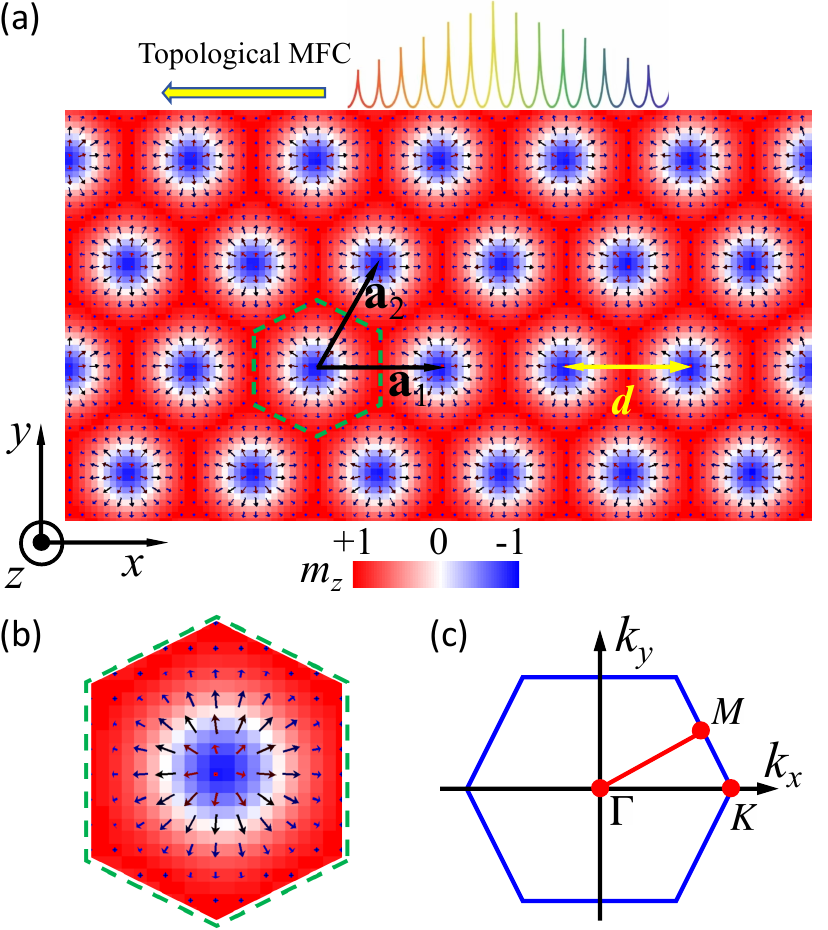}
\par\end{centering}
\caption{(a) Schematic of a two-dimensional triangular skyrmion lattice and the mechanism underlying topological magnon frequency comb generation. The out-of-plane magnetization component $m_z$ is color-coded. (b) Enlarged view of the unit cell encompassing a N\'eel-type skyrmion, as denoted in (a). (c) First Brillouin zone of the skyrmion lattice, with high-symmetry points $\Gamma$, $M$, and $K$ labeled.}
\label{Figure1}
\end{figure}

Very recently, Flower \textit{et al.} \cite{FlowerS2024} experimentally demonstrated topologically protected frequency combs in a photonic lattice, marking a seminal advance toward intrinsically robust comb generation. Leveraging analogies across (quasi)particles, the integration of magnetic topology and nonlinear effects harbors tremendous potential for realizing robust topological MFCs. Two-dimensional (2D) magnetic texture lattices, such as skyrmion lattices, serve as exemplary platforms for topological MFC explorations. In these systems, magnons undergo a Lorentz-like force induced by the emergent magnetic field from spin textures \cite{WeberS2022,MolinaNJP2016,DiazPRR2020,DiazPRL2019,ChenJAP2021}, resulting in nonzero Berry curvature that sustains topological magnon states \cite{XieJAP2021,WangJAP2020}.

In this Letter, we theoretically propose topological MFCs in a 2D triangular skyrmion lattice, without loss of generality, propelled by nonlinear interactions among topological magnons. We delineate their topological attributes via analytical models and micromagnetic simulations. Remarkably, these topological MFCs are orchestrated by four-magnon scattering mechanisms. Our insights spotlight the convergence of magnetic topology and nonlinearity, advancing the creation of topologically robust nonlinear magnonic architectures.

\textit{Model}---We examine a 2D triangular skyrmion lattice consisting of N\'eel-type spin textures, stabilized by interfacial Dzyaloshinskii-Moriya interaction (DMI) \cite{DzyaloshinskyJPCS1958,MoriyaPR1960}, as illustrated in Fig. \ref{Figure1}(a). The system's Hamiltonian includes exchange, DMI, anisotropy, and dipole-dipole terms
\begin{equation}\label{Eq1}
\begin{aligned}
\mathcal{H}=&-J\sum_{\langle i,j \rangle}\mathbf{S}_{i}\cdot\mathbf{S}_{j}-D\sum_{\langle i,j\rangle}(\hat{\mathbf{z}}\times \hat{\mathbf{r}}_{ij})\cdot(\mathbf{S}_{i}\times\mathbf{S}_{j})-K\sum_{i}(\mathbf{S}_{i}\cdot\hat{\mathbf{z}})^{2}\\&+\frac{\mu_{0}\mu_{s}^{2}}{4\pi S^{2}}\sum_{i<j}\frac{\mathbf{S}_{i}\cdot\mathbf{S}_{j}-3(\mathbf{S}_{i}\cdot\hat{\mathbf{r}}_{ij})(\mathbf{S}_{j}\cdot\hat{\mathbf{r}}_{ij})}{r_{ij}^{3}},
\end{aligned}
\end{equation}
where $\mathbf{S}_{i}\ (\mathbf{S}_{j})$ represents the spin angular momentum vector at site $i$ ($j$) with magnitude $|\mathbf{S}_{i}|=S$, $\langle i,j \rangle$ denotes summation over nearest-neighbor pairs, $\hat{\mathbf{r}}_{ij}=\mathbf{r}_{ij}/|\mathbf{r}_{ij}|$ is the unit vector from site $i$ to $j$ separated by distance $\mathbf{r}_{ij}$, $J$ is the exchange constant, $D$ is the DMI strength, $K$ is the uniaxial anisotropy constant (with easy axis along $z$), $\mu_{0}$ is the vacuum permeability, and $\mu_{s}$ is the atomic magnetic moment. Employing the Holstein-Primakoff (HP) transformation \cite{HolsteinPR1940}, we obtain the magnon bands of the skyrmion lattice by diagonalizing the magnon Hamiltonian (see Sec. A of Supplemental Material \cite{SM} for details).

The topological nature of the $j$-th magnon band is captured by the Chern number \cite{WangPRB2017,AvronPRL1983}
\begin{equation}\label{Eq2}
\mathcal{C}_{j}=\frac{1}{2\pi}\int_{\text{BZ}}d^{2}\mathbf{k}\mathcal{B}_{j}(\mathbf{k}),
\end{equation}
with the Berry curvature
\begin{equation}\label{Eq3}
\mathcal{B}_{j}(\mathbf{k})=i\text{Tr}\Bigg[\mathcal{P}_{j}(\mathbf{k})\Bigg(\frac{\partial \mathcal{P}_{j}(\mathbf{k})}{\partial k_{x}}\frac{\partial \mathcal{P}_{j}(\mathbf{k})}{\partial k_{y}}-\frac{\partial \mathcal{P}_{j}(\mathbf{k})}{\partial k_{y}}\frac{\partial \mathcal{P}_{j}(\mathbf{k})}{\partial k_{x}}\Bigg)\Bigg].
\end{equation}
The integral covers the full Brillouin zone (BZ), and $\mathcal{P}_{j}(\mathbf{k}) = \Psi_{j}(\mathbf{k})\eta\Gamma_{j}\Psi_{j}(\mathbf{k})^{\dag}\eta$ projects onto the $j$-th band \cite{WangJAP2021}. Here, $\Gamma_{j}$ is a diagonal matrix whose $j$-th diagonal entry is +1 and all other entries are zero, $\eta=\sigma_{z}\otimes I_{N}$, where $\sigma_{z}$ is the Pauli matrix and $I_{N}$ denotes the $N\times N$ identity matrix, and $\Psi_{j}(\mathbf{k})$ is the normalized eigenvector of the magnon Hamiltonian.

For theoretical computations, we adopt parameters emblematic of Co$|$Pt interfaces \cite{SampaioNN2013}: saturation magnetization $M_s = 0.58 \times 10^6$ A/m, exchange stiffness $A_{ex} = 1.5 \times 10^{-11}$ J/m, perpendicular magnetocrystalline anisotropy $K_u = 0.6 \times 10^6$ J/m$^3$, interfacial DMI $D_{ind} = 3$ mJ/m$^2$, and inter-skyrmion distance $d = 40$ nm [Fig. \ref{Figure1}(a)]. These map to the Hamiltonian parameters in Eq. \eqref{Eq1} as $J S^2 = 2 A_{ex} a$, $K S^2 = K_u a^3$, $D S^2 = D_{ind} a^2$, and $\mu_s = M_s a^3$ \cite{Rezendebook2020}. For seamless alignment with subsequent micromagnetic simulations, we choose lattice constant $a = 2$ nm, yielding a spin quantum number $S = M_s a^3 / (\gamma \hbar) = 250$, where $\gamma$ is the gyromagnetic ratio and $\hbar$ is the reduced Planck constant \cite{Rezendebook2020}. Dipole-dipole interactions are incorporated up to the fifth nearest neighbors \cite{AtxitiaPRB2010}.
\begin{figure}[t]
\begin{centering}
\includegraphics[width=0.49\textwidth]{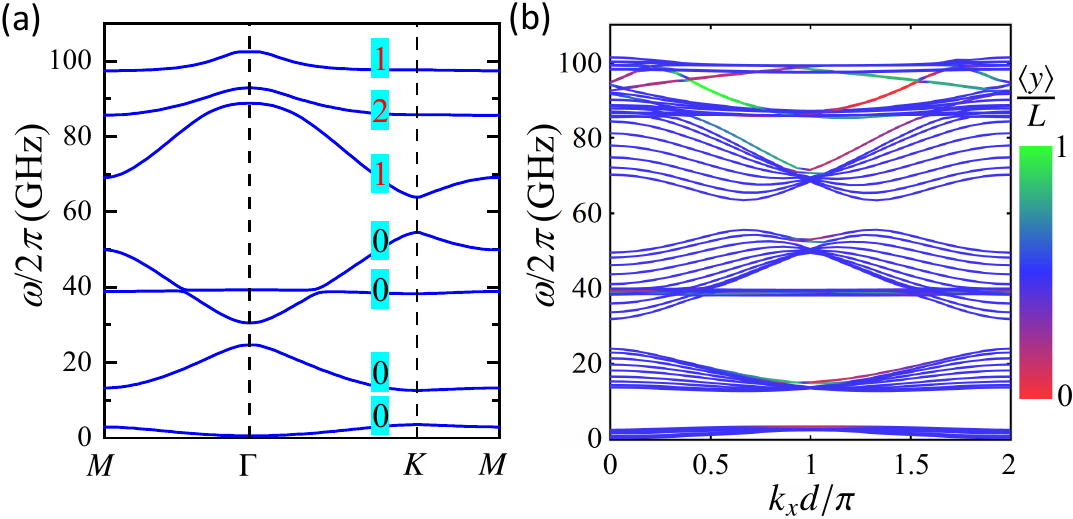}
\par\end{centering}
\caption{(a) Magnon band structure of an infinite 2D skyrmion lattice along the high-symmetry path $M - \Gamma - K - M$. Chern numbers for the lowest seven bands are labeled in cyan. (b) Magnon spectrum in a ribbon geometry with periodic boundary conditions along $x$ and finite width $L$ along $y$ encompassing 9 skyrmions. Modes are color-coded by average position $\langle y \rangle / L$, with greener (redder) shades indicating localization at the upper (lower) edge.}
\label{Figure2}
\end{figure}

\textit{Topological magnons in skyrmion lattices}---Figure \ref{Figure2}(a) illustrates the lowest seven magnon bands in an infinite 2D skyrmion lattice, with Chern numbers computed from Eq. \eqref{Eq2} labeled for each. The first four bands exhibit vanishing Chern numbers, whereas the ensuing three display nonzero values. According to the bulk-boundary correspondence \cite{HasanRMP2010,QiRMP2011}, bulk topological invariants dictate edge-state properties. In particular, for the $l$-th band gap, the winding number $\nu_l = \sum_{j < l} \mathcal{C}_j$ determines that $|\nu_l|$ corresponds to the number of topological edge states, with $\operatorname{sgn}(\nu_l)$ indicating their chirality \cite{MookPRB2014,HatsugaiPRL1993}.

To elucidate the topological magnon edge states, we calculate the magnon spectrum in a finite ribbon geometry, as shown in Fig. \ref{Figure2}(b). Periodic boundary conditions are applied along the $x$-direction, with the system featuring a finite width $L$ along $y$ that accommodates 9 skyrmions. Modes are color-coded according to their average position $\langle y \rangle / L$, with greener (redder) shades signifying localization at the upper (lower) edge. As apparent in Fig. \ref{Figure2}(b), three topological edge states emerge in the gap between the sixth and seventh bands, in accord with $\nu_6 = 3$. Moreover, given $\nu_5 = 1$ and $\nu_7 = 4$, chiral edge states are predicted in the fifth and seventh gaps. However, the fifth gap is exceptionally narrow, resulting in hybridization of edge modes with bulk states, and our analysis prioritizes modes below the seventh band. Hence, we center our attention on the edge states within the sixth band gap.

\begin{figure}[t]
\centering
\includegraphics[width=0.48\textwidth]{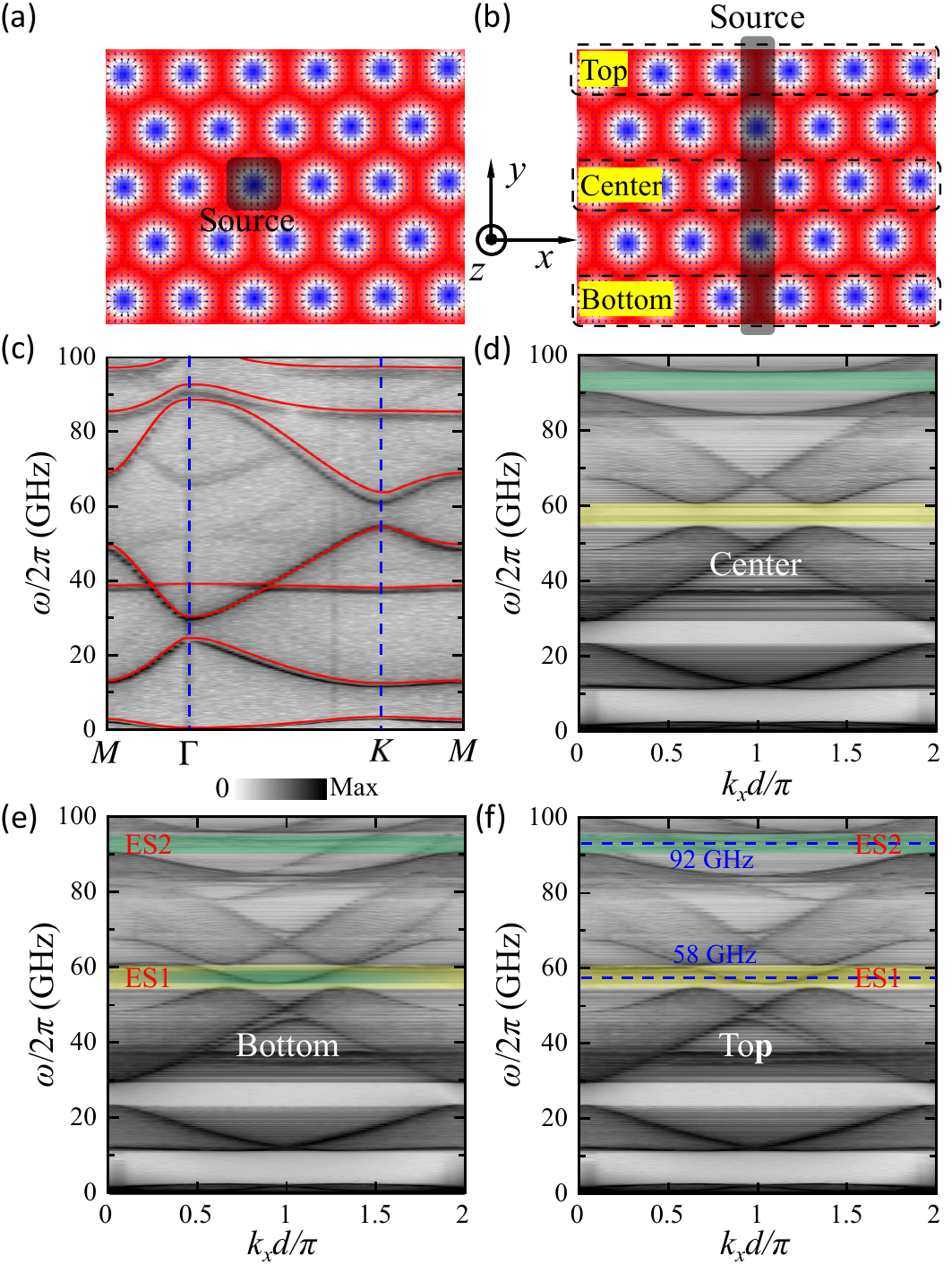}
\caption{(a) Schematic of the 2D skyrmion lattice with excitation field localized in a central quadrilateral region. (b) Excitation confined to a 40-nm-wide central nanostrip. (c) Magnon dispersion relations for an infinite skyrmion lattice; gray shading depicts micromagnetic simulations, red curves show theoretical results. Magnon spectra in the (d) central bulk, (e) bottom edge, and (f) top edge regions.}
\label{Figure3}
\end{figure}

To substantiate these theoretical insights, we perform micromagnetic simulations of the 2D skyrmion lattice using MuMax3 \cite{VansteenkisteAA2014}. The simulated specimen features lateral dimensions of $2 \times 2$
 $\mu$m$^2$ and a thickness of 2 nm, incorporating identical material parameters as in Fig. \ref{Figure2}. A modest Gilbert damping coefficient $\alpha = 0.001$ is employed to ensure well-resolved magnon dispersion relations, with a discretization cell size of $2 \times 2 \times 2$
 nm$^3$. For the magnon spectrum in an infinite 2D lattice, periodic boundary conditions are imposed along both $x$ and $y$ directions. Magnons are excited by a sinc-function magnetic field $\mathbf{H}(t) = H_0 \frac{\sin[2\pi f_c (t - t_0)]}{2\pi f_c (t - t_0)} \hat{\mathbf{x}}$, with $H_0 = 10$ mT, $f_c = 100$ GHz, and $t_0 = 1$ ns, applied within a central $40 \times 40$ nm$^2$ region [Fig. \ref{Figure3}(a)]. To probe edge magnon bands, open boundary conditions are adopted, with excitation restricted to a central 40-nm-wide nanostrip [Fig. \ref{Figure3}(b)].

Figure \ref{Figure3}(c) shows the magnon band structure below 100 GHz for an infinite 2D skyrmion lattice, where theoretical predictions (red curves) align closely with micromagnetic simulations (gray background). Under open boundaries, dispersion varies by region: the central bulk hosts multiple mode-free gaps [Fig. \ref{Figure3}(d)]. Strikingly, edge states, labeled ES1 and ES2, emerge at the bottom and top boundaries [Figs. \ref{Figure3}(e) and \ref{Figure3}(f)]. Analysis of their group velocities $d\omega/dk_x$ reveals that ES1 is bidirectional, while ES2 exhibits unidirectional propagation in the counterclockwise direction. The nonchiral ES1 stems from the Tamm-Shockley mechanism \cite{TammPZ1932,ShockleyPR1939}, whereby boundary-induced disruption of crystalline periodicity yields trivial localized states. Conversely, the chiral ES2 is topologically safeguarded by nonzero Chern numbers [Fig. \ref{Figure2}(a)]. We further verify the contrasting propagation behaviors of these nonchiral and chiral edge states (see Sec. B of Supplemental Material \cite{SM} for details).
\begin{figure}[t]
\centering
\includegraphics[width=0.48\textwidth]{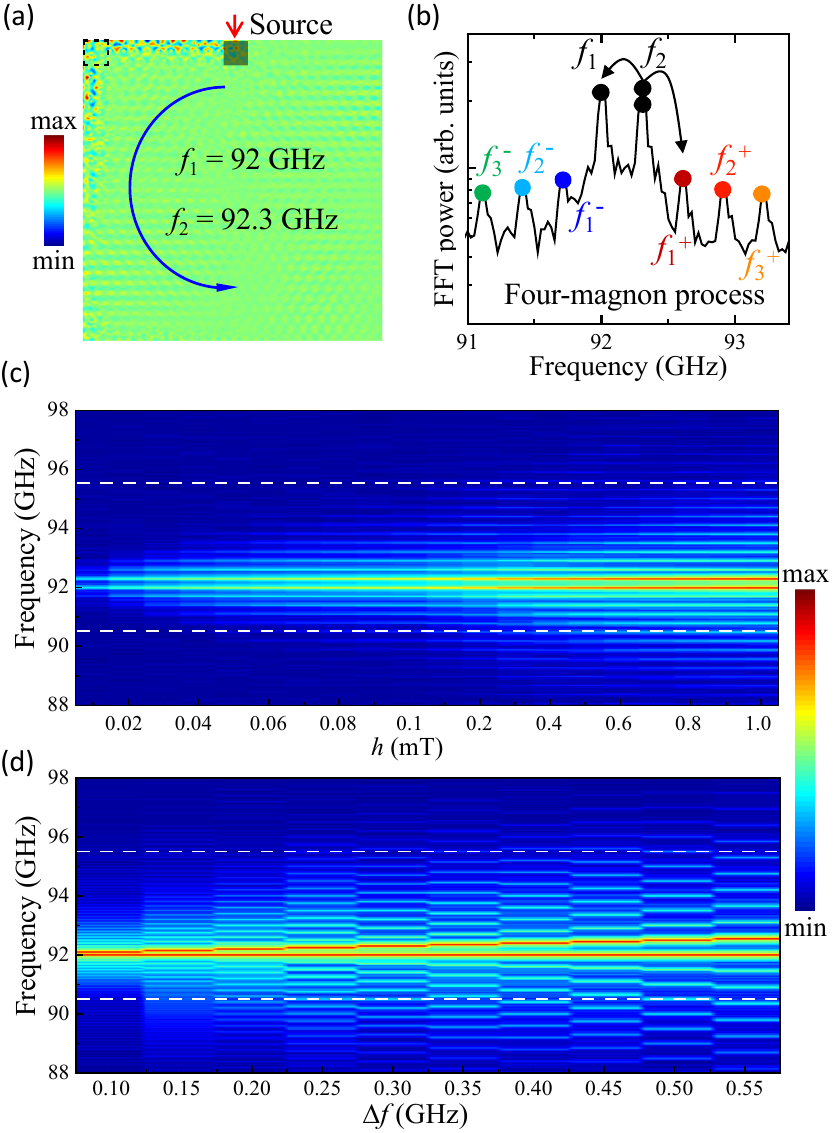}
\caption{(a) Magnon intensity snapshot under dual-frequency excitation at $f_1 = 92$ GHz and $f_2 = 92.3$ GHz, highlighting counterclockwise edge flow (blue arrow). (b) FFT spectrum and schematic of four-magnon scattering process $2f_2\rightarrow f_1+f_1^+$. (c) Magnon power spectra versus excitation amplitude $h$. (d) Spectra for varying detunings $\Delta f$ at $h = 1$ mT and $f_1 = 92$ GHz. Dashed white lines in (c) and (d) delimit the topological edge-state frequency range.}
\label{Figure4}
\end{figure}

\textit{Topological MFCs}---Prior studies have shown that MFCs arising from nonlinear interactions between magnons and skyrmion breathing modes typically feature comb spacings of tens of gigahertz \cite{WangPRL2021}. In our system, however, the topological magnon edge-state bandwidth is relatively narrow ($\sim$5 GHz; see Figs. \ref{Figure3}(e) and \ref{Figure3}(f)), hindering comb generation via traditional magnon-skyrmion scattering within this range. Instead, we exploit nonlinear couplings among the topological magnons themselves to realize topological MFCs.

A dual-frequency excitation field $\mathbf{h}(t) = h [\sin(2\pi f_1 t) + \sin(2\pi f_2 t)] \hat{\mathbf{x}}$ is applied at the source site [Fig. \ref{Figure4}(a)]. For concreteness, we choose $f_1 = 92$ GHz and $f_2 = 92.3$ GHz to explore nonlinear interactions among these topological magnon modes, with $\alpha = 0.01$ to suppress noise artifacts. Figure \ref{Figure4}(c) depicts the magnon spectra as a function of excitation amplitude $(h)$, obtained through fast Fourier transform (FFT) of the $y$-component of magnetization in the upper-left detection region [dashed black rectangle in Fig. \ref{Figure4}(a)]. Evidently, nonlinear couplings among topological magnons amplify with increasing $h$, producing progressively sharper MFCs with 0.3-GHz comb spacing. Furthermore, Fig. \ref{Figure4}(d) shows spectra for varying $f_2$ at fixed $h = 1$ mT and $f_1 = 92$ GHz, highlighting robust MFCs with readily tunable comb spacings. Fundamentally, this approach permits arbitrary comb spacings, contingent on appropriate damping and simulation duration. We ascertain that topological MFC formation is chiefly mediated by four-magnon scattering, not three-magnon processes (see Sec. C of Supplemental Material \cite{SM} for details). Figure \ref{Figure4}(b) schematizes this four-magnon process, wherein the frequencies of pumped magnons ($f_1$ and $f_2$) and scattered magnons ($f_1^+$, $f_1^-$, etc.) obey $f_1^+ = 2f_2 - f_1$ [see black arrows in Fig. \ref{Figure4}(b)], $f_1^- = 2f_1 - f_2$, $f_2^+ = f_1^+ + \Delta f$, $f_2^- = f_1^- - \Delta f$, $f_3^+ = f_2^+ + \Delta f$, and $f_3^- = f_2^- - \Delta f$, producing an MFC spaced by $\Delta f = f_2 - f_1$. For comparison, we also analyze MFCs in a uniformly magnetized film (see Sec. D \cite{SM}), where appropriate parameters can elicit MFCs but without unidirectional propagation, a key distinction from the topological MFCs reported here.

For analytical insight, we construct a Hamiltonian capturing four-magnon scattering
\begin{equation}\label{Eq4}
\begin{aligned}
\mathcal{H}_{4}=&\omega_1a_1^{\dag}a_1+\omega_2a_2^{\dag}a_2+\omega_{p}a_{p}^{\dag}a_{p}+g(a_{1}^{\dag}a_{p}^{\dag}a_2^{2}+{\rm H. c.})\\
&+h(a_1e^{i\omega_{1} t}+a_1^{\dag}e^{-i\omega_{1} t}+a_2e^{i\omega_{2} t}+a_2^{\dag}e^{-i\omega_{2} t}),
\end{aligned}
\end{equation}
where $a_1$ and $a_2$ denote the incident topological magnon modes at frequencies $\omega_1/2\pi= f_1$ and $\omega_2/2\pi= f_2$, respectively, and $a_p$ represents the scattered magnon mode at $\omega_p/2\pi =f_1^+$, satisfying $2 \omega_2 = \omega_1 + \omega_p$. The parameter $g$ quantifies the four-magnon coupling strength, while the final term describes the dual-frequency microwave drive with amplitude $h$ and frequencies $\omega_1$, $\omega_2$.

To partially remove time dependence in Eq. \eqref{Eq4}, we transition to a rotating frame via the unitary transformation $V_1 = \exp[-i \omega_1 t (a_1^\dagger a_1 + a_2^\dagger a_2 + a_p^\dagger a_p)]$. This yields the transformed Hamiltonian
\begin{equation}\label{Eq5}
\begin{aligned}
\mathcal{H}_{4}=&\Delta_2a_2^{\dag}a_2+\Delta_pa_p^{\dag}a_p+g(a_{1}^{\dag}a_p^{\dag}a_{2}^{2}+{\rm H. c.})\\&+h(a_1+a_2e^{i\Delta_2t}+{\rm H. c.}),
\end{aligned}
\end{equation}
where $\Delta_2 = \omega_2 - \omega_1$ and $\Delta_p = \omega_p - \omega_1$ denote the frequency detunings. From Eq. \eqref{Eq5}, the Heisenberg equations of motion, incorporating phenomenological damping, read
\begin{equation}\label{Eq6}
\begin{aligned}
i\frac{d a_1}{d t}&=g a_p^{\dag}a_2a_2+h-i\alpha_1 \omega_1 a_1,\\
i\frac{d a_2}{d t}&=\Delta_2 a_2+2g a_2^{\dag}a_pa_1+he^{-i\Delta_2t}-i\alpha_2 \omega_2 a_2,\\
i\frac{d a_p}{d t}&=\Delta_p a_p+g a_1^{\dagger}a_2a_2-i\alpha_p \omega_p a_p,\\
\end{aligned}
\end{equation}
with $\alpha_\mu$ ($\mu = 1, 2, p$) the mode-specific damping rates. Equation \eqref{Eq6} admits steady-state solutions of the form $a_1(t) = \bar{a}_1$, $a_2(t) = \bar{a}_2 e^{-i \Delta_2 t}$, and $a_p(t) = \bar{a}_p e^{-2 i \Delta_2 t}$. Assuming equal damping rates $\alpha_1 = \alpha_2 = \alpha_p = \alpha$ (the Gilbert damping \cite{GilbertIEEE2004}) and weak nonlinearity, solutions approximate $\bar{a}_1 \approx h / (i \alpha \omega_1)$, $\bar{a}_2 \approx h / (i \alpha \omega_2)$, and $\bar{a}_p \approx g h^3 / [i \alpha^3 \omega_2^2 \omega_1 (\Delta_p + i \alpha \omega_p)]$. Clearly, dual-frequency excitation triggers four-magnon processes sans amplitude threshold, unlike three-magnon analogs \cite{WangPRL2021,WangPRL2022}. Micromagnetic simulations [Fig. \ref{Figure4}(c)] affirm this, with MFCs evident at fields down to $0.01$ mT.

To validate the topological essence of MFCs in Figs. \ref{Figure4}(c) and \ref{Figure4}(d), we inspect the spatial intensity distribution under dual-frequency driving at $f_1 = 92$ GHz, $f_2 = 92.3$ GHz, and $h = 1$ mT [Fig. \ref{Figure4}(a)]. The generated signals conspicuously localize to boundaries and circulate counterclockwise. These edge modes further demonstrate robust propagation, undeterred by backscattering into the bulk at sharp 90$^\circ$ corners [dashed black rectangle in Fig. \ref{Figure4}(a)]. Such traits confirm that comb teeth stem from topological edge states, preserving protection during nonlinear dynamics and thus embodying topological MFCs.

\textit{Discussion and conclusion}---While conventional MFCs have been theoretically and experimentally realized in diverse magnetic systems, topological MFCs, where comb generation is intrinsically protected by topological invariants, have remained unexplored in magnonics. Our study inaugurates this field by proposing topological MFCs via nonlinear four-magnon scattering purely among topological magnons in a two-dimensional triangular skyrmion lattice \cite{MuhlbauerS2009,YuN2009}. This produces robust combs localized to chiral edge states, with tunable spacings and absent amplitude threshold, starkly differing from prior MFCs dependent on defect-vulnerable trivial magnons or hybrid magnon-soliton/phonon couplings. In contrast to anisotropic or nonreciprocal MFCs in antiferromagnets or magnon-phonon hybrids, our Berry curvature-based strategy shields comb emergence and transport from imperfections, heralding a revolutionary advance in fault-tolerant nonlinear magnonics. We envisage rapid experimental realization of topological MFCs in MnSi skyrmion lattices \cite{WeberS2022}, potentially via microwave spectroscopy techniques, Brillouin light scattering, and spin pumping methods.

In conclusion, we theoretically demonstrate nonlinear couplings among topological magnons in a two-dimensional skyrmion lattice, thereby introducing topological MFCs, an exotic paradigm in magnonics. Unlike conventional MFCs predicated on defect-prone trivial magnons, these topological variants emerge resiliently through dual-frequency driving within topological bands at modest amplitudes, featuring seamlessly tunable comb spacings via frequency detuning and no inherent threshold. Our results furnish a blueprint for defect-resilient, topologically protected MFC devices with exceptional robustness, primed to transform spintronic technologies in high-precision information processing, quantum sensing, and computing. Furthermore, by bridging magnetic topology and nonlinearity, this work opens expansive horizons for future inquiries, encompassing experimental validations in varied skyrmionic platforms, generalizations to alternative topological structures (e.g., merons or hopfions), synergies with non-Hermitian magnonics or quantum hybrids, and the unveiling of unprecedented phenomena at the topology-nonlinearity crossroads.
\begin{acknowledgments}
\textit{Acknowledgments}---We thank Z. Wang and X. S. Wang for helpful discussions. This work was supported by the National Key R$\&$D Program under Contract No. 2022YFA1402802, the National Natural Science Foundation of China (NSFC; Grants No. T2495212, No. 12074057, and No. 12374103), and Sichuan Science and Technology Program (No. 2025NSFJQ0045). Z.-X.L. acknowledges financial support from the Natural Science Foundation of Hunan Province of China (Grant No. 2023JJ40694). X.-J.L. acknowledges support from the Talent Introduction Program of Chengdu Normal University under Grant No.YJRC2021-14. Z. J. acknowledges the financial support from NSFC (Grant No. 12404125) and the China Postdoctoral
Science Foundation (Grant No. 2024M750337). X.-G.Z. was supported by the NSFC (Grant No. 52271239) and the Guangdong Basic and Applied Basic Research
Foundation (Grant No. 2022B1515120058).  
\end{acknowledgments}

Z. Li, X. Liu, and Z. Jin contributed equally.

\end{document}